\documentclass[11pt,a4paper]{article} 
\usepackage{jcappub}
\usepackage{graphicx}
\usepackage{subfigure}
\usepackage{dcolumn}
\usepackage{bm}

\title{Indirect dark matter detection in the light of sterile neutrinos}

\author[a]{Arman Esmaili}
\author[a,b,c]{, O. L. G. Peres}
\emailAdd{aesmaili@ifi.unicamp.br}
\emailAdd{orlando@ifi.unicamp.br}
\affiliation[a]{Instituto de Fisica Gleb Wataghin - UNICAMP, 13083-859, Campinas, SP, Brazil}
\affiliation[b]{Arizona State University, Tempe, AZ 85287-1504}
\affiliation[c]{The Abdus Salam International Centre for Theoretical Physics\\
Strada Costiera 11, I-34014 Trieste, Italy}

\abstract{The recent global fit of short baseline neutrino oscillation data favors the presence of one (or more) sterile neutrino state which leads to new mass splitting $\Delta m^2\sim 1\,{\rm eV}^2$. We consider the effect of this new states on the evolution of neutrinos from the dark matter annihilation inside the Sun. We show that neutrinos with energy $E_\nu\gtrsim100\,{\rm GeV}$ undergo resonant active-sterile oscillation which depletes the flux of neutrinos arriving at the Earth. As an example of this effect, we present the oscillation probabilities for the case of monochromatic neutrinos from the direct annihilation of dark matter particles to neutrinos and the depletion due to the presence of sterile neutrinos. We discuss the seasonal variation of oscillation probabilities which is expected for the case of monochromatic neutrinos.}


\begin{document}

\maketitle

\section{\label{sec:intro}Introduction}

Well-established evidences ranging from galaxy to clusters of galaxies and cosmological scales are in favor of the existence of a huge amount of matter in the universe, the so-called Dark Matter (DM) \cite{dm}. In spite of the compelling evidences on the existence of DM, almost nothing is known about the nature of DM and especially the fundamental particle(s) that can comprise it. Basically two approaches are used in experiments aiming at the detection of DM: direct and indirect methods. The direct detection method is based on measurement of the scattering of particles comprising DM (hereafter $\chi$) off the nuclei of ordinary matter. The current upper bounds, from experiments aiming at direct detection, on the spin-independent $\sigma_{\chi N}^{{\rm SI}}$ and spin-dependent $\sigma_{\chi N}^{{\rm SD}}$ cross sections are respectively  $\sim10^{-8} \, {\rm pb}$ \cite{xenon100} and $\sim10^{-2} \,{\rm pb}$ \cite{simplesd,couppsd} for DM mass $m_\chi\sim100\, {\rm GeV}$.    

The indirect method is based on the detection of annihilation products of DM particles including gamma rays, charged particles and neutrinos. To attain a substantial rate of $\chi\bar{\chi}$ (where $\bar{\chi}=\chi$ in the case of a Majorana DM particle) annihilation, the density of DM should be sizable which is the case in the center of the galaxies and astrophysical objects, and the closest one to us is the Sun which we consider in this paper. The DM particles which scatter elastically off the nuclei in Sun lose a part of their momentum and if their velocity drop below the escape velocity in the Sun, they trap gravitationally and accumulate in the center of the Sun ({\it see} \cite{jungman} and references therein). Accumulation of the DM particles trapped by the gravitational potential well of the Sun increases the number density $n_{\chi}$ and since the annihilation rate $\Gamma \propto n_{\chi}^2$, large values of $n_{\chi}$ can result in sizable annihilation rate.  

Annihilation of DM particles inside the Sun can produce neutrinos in two ways: 1) As the secondary product of the annihilation modes  $\chi\bar{\chi}\to q\bar{q}$, fermion pairs, gauge bosons or gauge bosons and Higgs particle. In this case the neutrino beam has a continuous energy spectrum that its shape depends on the annihilation mode;  2)  As the direct annihilation of the DM particles $\chi\bar{\chi}\to \nu\bar{\nu}$. In this case the neutrino beam is monochromatic with energy equal to the DM particle's mass. It should be noted that although the branching ratio of the annihilation mode $\chi\bar{\chi}\to \nu\bar{\nu}$ is predicted to be small in some models (especially in MSSM where the annihilation cross section of the lightest neutralino to neutrinos is suppressed by the factor $m_{\nu}^2/m_{\chi}^2$), it is possible to build models that this suppression is circumvented ({\it see} \cite{farzan2,lindner} for such models). It can be shown that if the DM-nuclei cross section saturates the upper bound from the direct method experiments, the number density $n_\chi$ is large enough to produce detectable flux of neutrinos in neutrino telescopes such as IceCube \cite{farzan1}.

The recent global fit of short baseline neutrino oscillation data, including LSND \cite{lsnd}, MiniBooNE \cite{miniboone} and reactor anomaly \cite{reactor}, shows indication in favor of active-sterile oscillation with mass splitting $\Delta m^2\sim 1\,{\rm eV}^2$ \cite{giunti}. In this paper we consider the effect of this new mass splitting on the evolution of neutrino beam from DM annihilation inside the Sun. We show that the presence of a mass splitting $\Delta m^2\sim 1\,{\rm eV}^2$ would lead to resonant flavor conversion for neutrinos with energy $E_\nu\gtrsim 100\,{\rm GeV}$; similar to the resonant flavor conversion of thermonuclear solar neutrinos with $E_\nu\sim{\rm MeV}$ and $\Delta m_{\rm sol}^2\sim 8\times 10^{-5}\,{\rm eV}^2$. We show that due to this resonant flavor conversion, the electron neutrinos and muon/tau anti-neutrinos produced in the annihilation of DM particles with mass $m_\chi\gtrsim 100\,{\rm GeV}$ are mostly sterile when they leave the Sun; which results in a reduction of the expected number of events in neutrino telescopes such as IceCube \cite{icecube}.

As an example of the consequences of this resonant flavor conversion, we consider the case of monochromatic neutrinos from the annihilation $\chi\bar{\chi}\to\nu\bar{\nu}$. It was proposed in \cite{farzan1,farzan3} that in the case of monochromatic neutrinos the number of events at neutrino telescopes on the Earth (such as IceCube) would show a seasonal variation due to the change in the Earth-Sun distance during the year. We calculate the flavor oscillation probabilities from Sun to the IceCube site for monochromatic neutrinos and the expected deviations in the seasonal variation due to the presence of $\Delta m^2\sim 1\,{\rm eV}^2$. The analysis for the case of secondary neutrinos with continuous energy spectrum from the annihilations $\chi\bar{\chi}\to q\bar{q},W^+W^-,\ldots$ will appear in \cite{future}.

This paper is organized as follows. In section~\ref{sec:sun} we discuss the flavor oscillation of neutrinos inside the Sun and the conditions that a resonant flavor oscillation would take place. In section~\ref{sec:earth} we calculate the oscillation probabilities at the Earth and we discuss the deviations in the seasonal variation resulting from the active-sterile oscillation. The conclusions are given in section~\ref{sec:conc}.

\section{\label{sec:sun}Oscillation inside the Sun}

The presence of new sterile neutrino states with mass splitting $\Delta m^2\sim 1\,{\rm eV}^2$ favored by the global analysis of short baseline neutrino oscillation data would change the evolution pattern of neutrinos inside the Sun. Especially, the matter effect of the Sun's medium would induce a MSW resonance \cite{msw} which can drastically change the flavor content of neutrino beam at the surface of the Sun. In this paper we assume one new mass eigenstate $\nu_4$ (the so-called 3+1 scenario \cite{31}) with mass splittings $\Delta m_{41}^2 \simeq \Delta m_{42}^2 \simeq \Delta m_{43}^2 \sim 1\,{\rm eV}^2$, where $\Delta m_{ij}^2\equiv m_i^2-m_j^2$. To calculate the flavor oscillation probabilities we use the following convention for the mixing matrix (assuming all the CP-violating phases equal to zero):
\begin{equation}\label{u}
U = R_{34} R_{24} R_{14} R_{23} R_{13} R_{12} 
\end{equation}
where $R_{ij}$ (with $i,j=1,2,3,4$ and $i<j$) denotes the 4-dimensional rotation matrix in $i$-$j$ plane with the rotation angle $\theta_{ij}$ (we use the notation $s_{ij}\equiv\sin\theta_{ij}$ and $c_{ij}\equiv\cos\theta_{ij}$). For the conventional mixing angles $\theta_{12}$, $\theta_{23}$ and $\theta_{13}$ we use the values from \cite{schwetz}. The current best-fit values for $\theta_{14}$ and $\theta_{24}$ from the global analysis of short baseline neutrino oscillation data \cite{giunti} are: $s^2_{14}=0.03$ and $s^2_{24}=0.012$. For the mixing angle $\theta_{34}$ the strongest limits comes from the MINOS experiment \cite{minos}: $s^2_{34}\leq0.09$.

The evolution equations of the flavor neutrino states inside the Sun are
\begin{equation}
i\frac{{\rm d} \nu_\alpha}{{\rm d}r}= \left[\frac{1}{2E_\nu} \left( U\bm{M}^2U^\dagger \right)+ \bm{V}(r)\right]_{\alpha\beta}\nu_\beta~,
\end{equation}
where $r$ is the distance from Sun's center; $\alpha,\beta=e,\mu,\tau,s$; and 
$$\bm{M}^2={\rm diag}\,(0,\Delta m_{{21}}^2,\Delta m_{{31}}^2,\Delta m_{41}^2)~,$$ 
where $\Delta m_{21}^2\sim8\times10^{-5}\,{\rm eV}^2$ and $\Delta m_{31}^2\sim2\times10^{-3}\,{\rm eV}^2$ \cite{schwetz}; and
$$\bm{V}(r)=\sqrt{2}\,G_F\, {\rm diag}\,(N_e(r),0,0,N_n(r)/2).$$
$N_e$ and $N_n$ are respectively the electron and neutron number density profiles of the Sun.

In the two flavors approximation, the MSW \cite{msw} resonance energy for the active-sterile oscillation is given by
\begin{equation} \label{reses}
E_{{\rm res}}=6.55\times10^3\, {\rm GeV}\left(\frac{\Delta m_{41}^2}{1\,{\rm eV}^2} \right) \left(\frac{\cos 2\theta}{1} \right) \left(\frac{N_A {\rm cm}^{-3}}{N_{eff}} \right) ,
\end{equation} 
where $N_A$ is the Avogadro's number and $\theta$ is $\theta_{14}$, $\theta_{24}$ and $\theta_{34}$ for $e-s$, $\mu-s$ and $\tau-s$ resonance, respectively. However, due to the hierarchical mass scheme in 3+1 scenario and the experimental bounds  on the active-sterile mixing angles, the two flavors approximation for the resonance energy is quite accurate. It should be emphasized that although we use eq.~(\ref{reses}) for approximating the resonance energy, we calculate the probabilities in the realistic four flavors scheme numerically. For $e-s$ resonance: $N_{eff}=N_e-N_n/2$ and for $(\mu/\tau)-s$ resonance: $N_{eff}=N_n/2$. Also, we assume normal hierarchy for the active-sterile mass splitting; {\it i.e.}, $\Delta m_{41}^2\geq0$. It should be noticed that the $e-s$ resonance occurs for neutrinos and $(\mu/\tau)-s$ resonance occurs for the anti-neutrinos. The minimum value of the resonance energy, for fixed values of $\Delta m_{41}^2$ and $\cos 2\theta$, corresponds to the maximum value of $N_{eff}$ which is at the center of Sun. According to Standard Solar Model \cite{ssm}, at the center of Sun we have
\begin{equation*}
{\rm Max}\left[\left(N_{eff}\right)^{e-s}\right]= \left.\left(N_e-\frac{N_n}{2}\right)\right|_{{\rm center}}=76.9 \quad\frac{N_A}{{\rm cm}^3}~,
\end{equation*}
\begin{equation*}
{\rm Max}\left[\left(N_{eff}\right)^{\mu/\tau-s}\right]=\left.\left(\frac{N_n}{2}\right)\right|_{{\rm center}}=26.9 \quad\frac{N_A}{{\rm cm}^3}~.
\end{equation*}
Thus, it is easy to see from eq.~(\ref{reses}) that the evidences from short baseline neutrino experiments on the existence of a new sterile state with $\Delta m_{41}^2 \simeq 1\, {\rm eV}^2$ and small mixing angles $\cos 2\theta_{i4}\simeq1$ ($i=1,2,3$), would lead to resonant flavor conversion $\nu_e\to\nu_s$ for electron neutrinos with energy $E_\nu\gtrsim85\,{\rm GeV}$ and $\bar{\nu}_{\mu/\tau}\to\bar{\nu}_s$ for muon and tau anti-neutrinos with energy $E_\nu\gtrsim 240\,{\rm GeV}$. 

\begin{figure}[t]
\center
\includegraphics[scale=0.6]{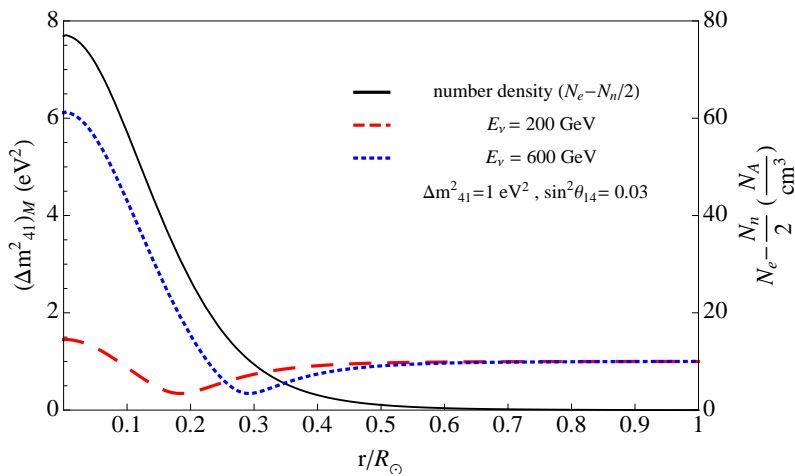}
\caption{\label{mass} The effective mass-squared difference in matter $(\Delta m_{41}^2)_M$ (left vertical axis) and $N_e-N_n/2$ (right vertical axis) with respect to the distance from Sun's center. The red (dashed) and blue (dotted) curves correspond respectively to neutrinos with energy $E_\nu=200\, {\rm GeV}$ and 600~GeV at the center of Sun and the black (solid) curve shows the number density. In this plot we assumed: $\sin^2 \theta_{14}=0.03$ and $\Delta m_{41}^2=1\,{\rm eV}^2$. }
\end{figure}

To illustrate the resonance behavior we plot in figure~\ref{mass} the effective mass-squared difference in matter $(\Delta m_{41}^2)_M$ with respect to the radial distance from the center of Sun. In this plot we assumed: $s^2_{14}=0.03$ and $\Delta m_{41}^2=1\,{\rm eV}^2$. The red (dashed) and blue (dotted) curves correspond respectively to neutrinos with energy $E_\nu=200\, {\rm GeV}$ and 600~GeV. The minima in these curves show the resonance point in the trajectory of electron neutrinos toward the surface of Sun. At the same plot of figure~\ref{mass}, the black (solid) curve shows the number density of $N_e-N_n/2$ (with the scale at the right vertical axis) which plays role in the $e-s$ resonant flavor oscillation. As can be seen the resonance occur at the distance from the Sun's center where the condition in eq.~(\ref{reses}) is satisfied. The DM particles trapped inside the Sun, would virialize with the Sun's medium. Comparing the gravitational potential energy of the Sun's matter with the average thermal energy of DM it can be shown that DM particles reside in the center of Sun in a radius $r_{\rm DM}\lesssim 0.01 R_\odot$. On the other hand, according to the Standard Solar Model, the matter density in the Sun is almost constant within the radius $0.05 R_\odot$, which means that all the neutrinos produced from $\chi\bar{\chi}$ annihilation with energy higher than the resonance energy in eq.~(\ref{reses}) would undergo MSW active-sterile resonance.

\begin{figure}[t]
\center
\includegraphics[scale=0.6]{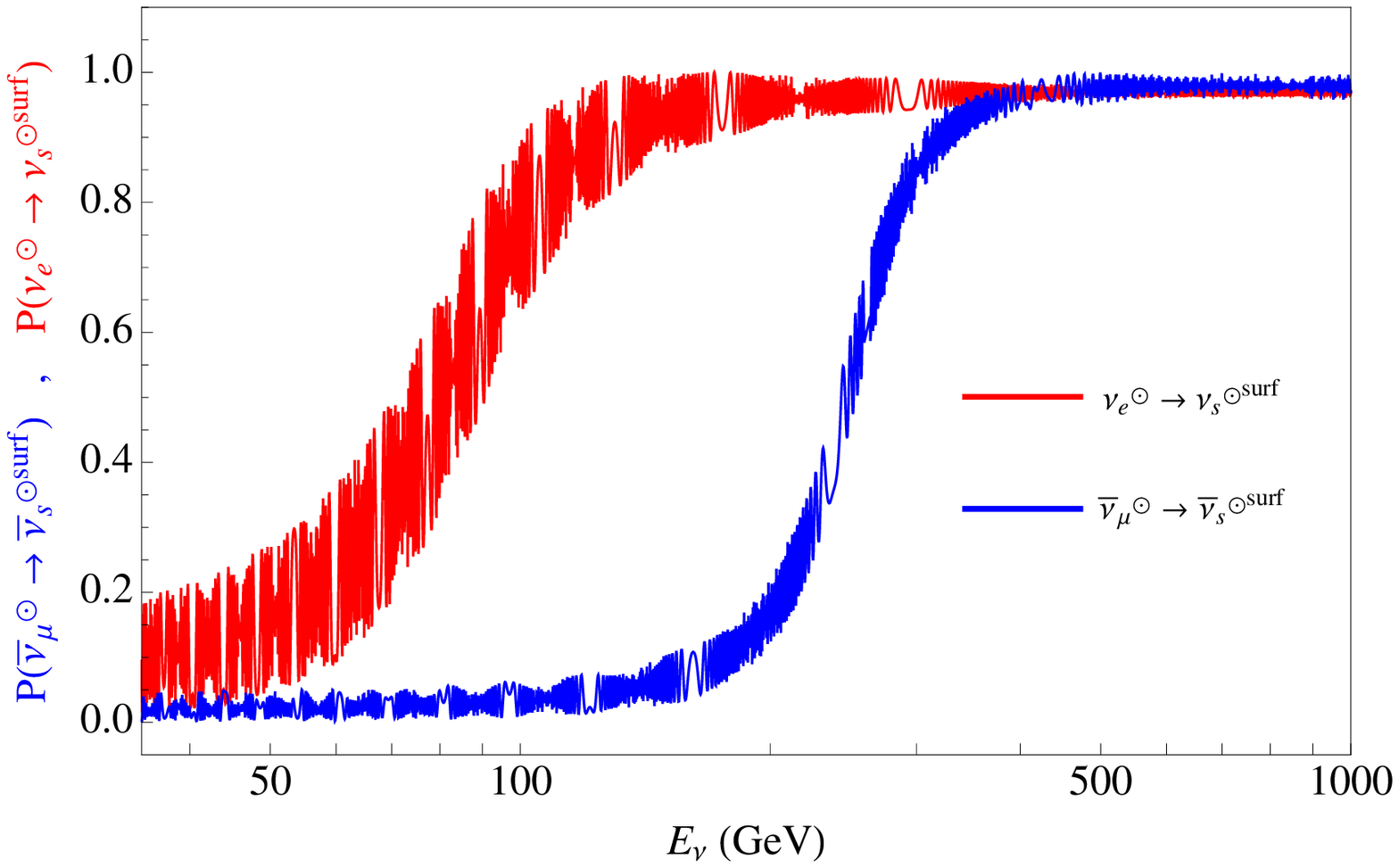}
\caption{\label{probsun1} The probabilities $P(\nu_e^\odot\to\nu_s^{\odot\rm surf})$ (red curve) and $P(\bar{\nu}_{\mu}^\odot\to\bar{\nu}_s^{\odot\rm surf})$ (blue curve) with respect to the neutrino energy. In this plot we assumed  ($\sin^2 \theta_{14}=0.03,\theta_{24}=0,\theta_{34}=0$) and ($\theta_{14}=0,\sin^2 \theta_{24}=0.01,\theta_{34}=0$) for $\nu_e^\odot\to\nu_s^{\odot\rm surf}$ and $\nu_\mu^\odot\to\nu_s^{\odot\rm surf}$, respectively; and for the both cases $\Delta m_{41}^2=1\,{\rm eV}^2$. }
\end{figure}

We plot in figure~\ref{probsun1} the probabilities $P(\nu_e^\odot\to\nu_s^{\odot\rm surf})$ (red curve) and $P(\bar{\nu}_{\mu}^\odot\to\bar{\nu}_s^{\odot\rm surf})$ (blue curve) with respect to neutrino energy; where by $\nu_\alpha^\odot$ and $\nu_\alpha^{\odot\rm surf}$ we mean respectively the neutrino state at the Sun's center and surface. In this plot we assumed  ($s^2_{14}=0.03,\theta_{24}=0,\theta_{34}=0$) for $\nu_e^\odot\to\nu_s^{\odot\rm surf}$; and ($\theta_{14}=0,s^2_{24}=0.01,\theta_{34}=0$) for $\bar{\nu}_\mu^\odot\to\bar{\nu}_s^{\odot\rm surf}$; and $\Delta m_{41}^2=1\,{\rm eV}^2$ for the both curves. As can be seen from the curves, the onset of the active-sterile flavor conversion occurs at the energy given by eq.~(\ref{reses}), that is $\sim 85 \,{\rm GeV}$ for $\nu_e$ and $\sim 240 \,{\rm GeV}$ for $\bar{\nu}_{\mu/\tau}$. The rapid oscillation in both of the curves in figure~\ref{probsun1} are due to the 1-4 mixing. The vacuum oscillation length of 1-4 mixing is given by
\begin{equation}\label{14osc}
L_{{\rm osc}}^{41}=\frac{4\pi E_\nu}{\Delta m_{41}^2}=246 \, {\rm km} \left(\frac{E_\nu}{100\, {\rm GeV}} \right) \left( \frac{1 \, {\rm eV}^2}{\Delta m_{41}^2} \right)~.
\end{equation}
The 14-induced oscillation contribute the two following terms: $\sin^2 2\theta_{14}\sin^2(\pi R_\odot /L_{{\rm osc}}^{41})$ and $\sin^2 2\theta_{24}\sin^2(\pi R_\odot /L_{{\rm osc}}^{41})$ to the $\nu_e^\odot\to\nu_s^{\odot\rm surf}$ and $\bar{\nu}_\mu^\odot\to\bar{\nu}_s^{\odot\rm surf}$ oscillations, respectively; where $R_\odot\simeq 7\times 10^5 \, {\rm km}$ is the radius of Sun. In figure~\ref{probsun1} we assumed $\sin^2 2\theta_{14}=0.11$ and $\sin^2 2\theta_{24}=0.04$; which justifies the relative smaller amplitude of rapid oscillation in the $\bar{\nu}_\mu^\odot\to\bar{\nu}_s^{\odot\rm surf}$ curve with respect to the $\nu_e^\odot\to\nu_s^{\odot\rm surf}$ curve.

\begin{figure}[t]
\centering
\centerline{\includegraphics[scale=0.45]{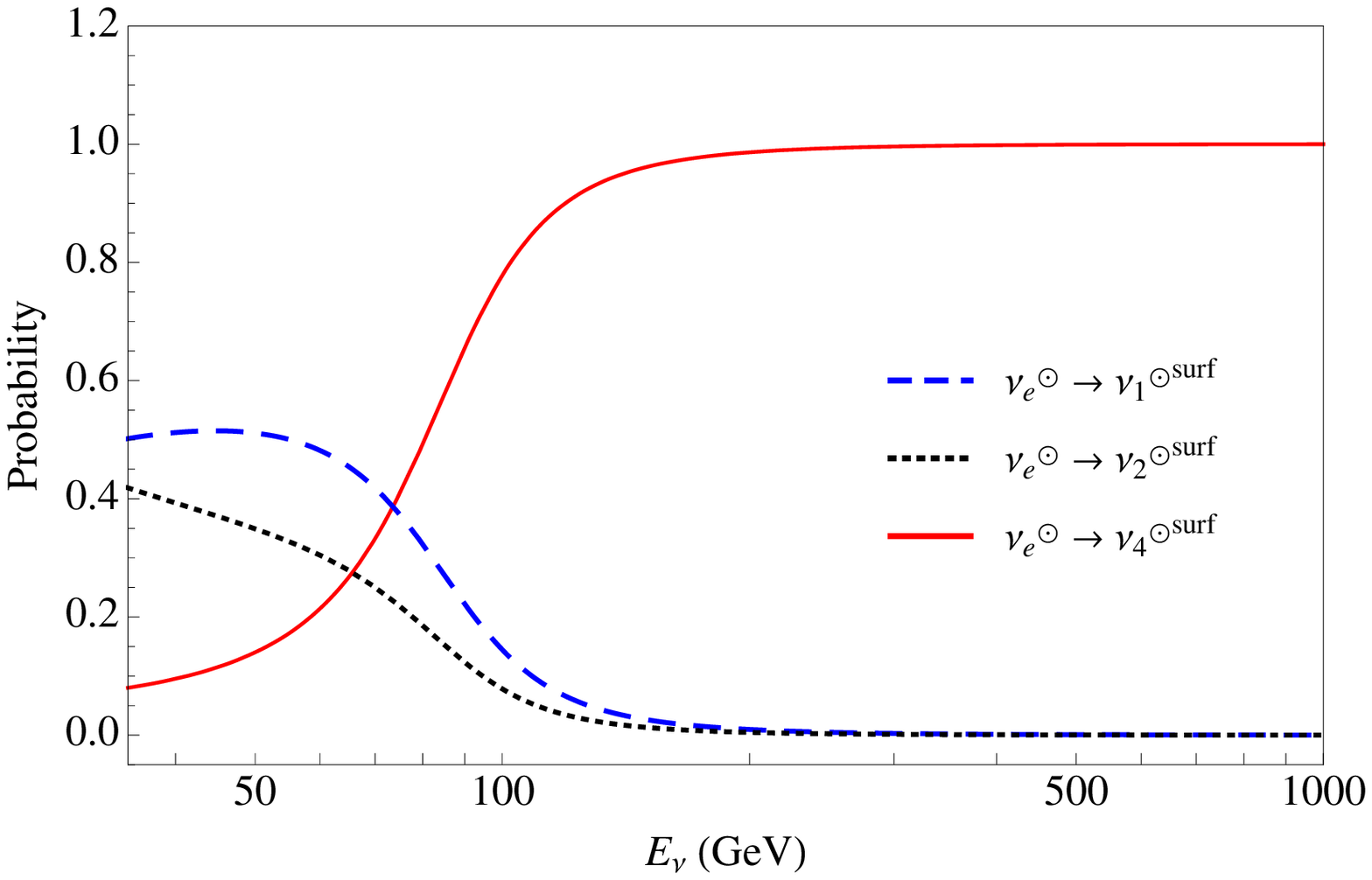}
  \includegraphics[scale=0.45]{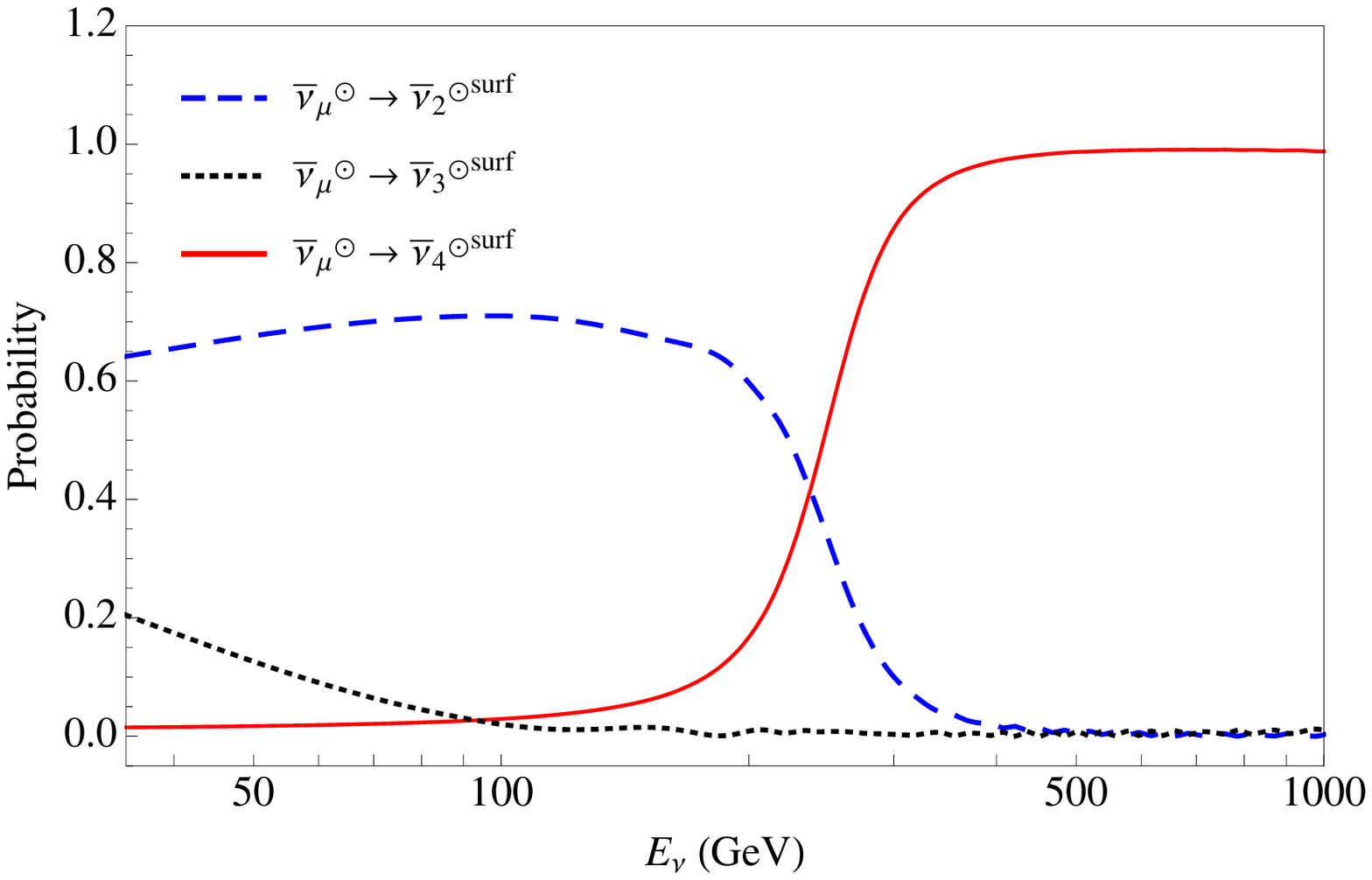}}
\caption{\label{flavortomass} The probabilities $P(\nu_e^\odot\to\nu_i^{\odot\rm surf})$ (left) and $P(\bar{\nu}_\mu^\odot\to\bar{\nu}_i^{\odot\rm surf})$ (right) . For the left plot we assumed ($s^2_{14}=0.03,s^2_{24}=s^2_{34}=0$) and for the right plot ($s^2_{14}=0,s^2_{24}=0.01,s^2_{34}=0$); and for the both plots $\Delta m_{41}^2=1\,{\rm eV}^2$.  }
\end{figure}

For the case of electron neutrino at the center of Sun, $\nu_e^\odot$, with energy $E_\nu\gtrsim85\,{\rm GeV}$, the resonant flavor oscillation due to $\Delta m_{41}^2$ plays role such that the state of the neutrino at the surface of Sun is almost $\nu_4^{\odot\rm surf}$. The left plot in figure~\ref{flavortomass} shows the numerical calculation of probabilities $P(\nu_e^\odot\to\nu_i^{\odot\rm surf})$ with respect to the neutrino energy for the mixing parameters the same as for red curve in figure~\ref{probsun1}. The right plot in figure~\ref{flavortomass} shows the probabilities for the case of $\bar{\nu}_\mu^\odot$ at the center of Sun and again with the same mixing parameters as the blue curve in figure~\ref{probsun1}. For the $\bar{\nu}_\mu^\odot$ also the MSW resonance makes the state of neutrino almost $\bar{\nu}_4^{\odot\rm surf}$ for energies $E_\nu\gtrsim240\,{\rm GeV}$, as can be seen from the plot. However it should be noticed that the complete conversion $\bar{\nu}_\mu^\odot\to\bar{\nu}_4^{\odot\rm surf}$ occurs when $s_{34}=0$. Assuming $s_{34}\neq0$, the state of the neutrino at the Sun's surface is the superposition of $\bar{\nu}_4^{\odot\rm surf}$ and $\bar{\nu}_2^{\odot\rm surf}$.

From the surface of Sun to the surface of Earth neutrinos evolve in vacuum. For a continuous energy spectrum which is the case for secondary neutrinos produced in the annihilation modes $\chi\bar{\chi}\to$($q\bar{q}$, fermions, gauge bosons), all the oscillatory terms between Sun and Earth average out and the oscillation probabilities are the same as at the surface of Sun. However, for the monochromatic neutrinos from the annihilation modes $\chi\bar{\chi}\to\nu\bar{\nu}$ the oscillatory terms can play role which we will discuss in the next section.

An important feature of this resonant conversion is that it take place even for very small values of mixing angle $\theta_{14}$, $\theta_{24}$ and $\theta_{34}$.  Considering the adiabaticity of conversion, it can be shown that resonance occur for $(s_{14}^2,s_{24}^2,s_{34}^2)\gtrsim10^{-3}$, which is far below the present upper limits from the short baseline experiments. Thus, we propose that in the indirect DM searches in the neutrino telescopes such as IceCube, this point should be considered. In the next section we discuss this point for the case of monochromatic neutrinos in more detail.  

Recently the paper~\cite{Arguelles:2012cf} appeared that discuss some of the topics discussed here. In \cite{Arguelles:2012cf} the authors consider the same effect as discussed in this paper and also they derive the upper limit on the DM-nucleon cross section in the presence of sterile neutrinos. However, in this paper we consider the case of monochromatic neutrinos and we discuss the seasonal variation effect which is expected for monochromatic neutrinos.

\section{\label{sec:earth}Oscillation probability at the Earth for monochromatic neutrinos}

The neutrinos produced in $\chi\bar{\chi}\to \nu\bar{\nu}$ are monochromatic with the energy $E_{\nu}=m_\chi$ and thus the spectrum of neutrinos at the production point is a sharp line.  It has been shown in \cite{farzan3} that thermal velocity distribution of $\chi$, natural wave-packet width of neutrinos at the production point and decoherence due to gravitational acceleration of $\chi$ do not smear the sharp line feature of the spectrum. The other two potential sources of smearing are neutral current interaction of neutrinos with the nuclei in Sun's medium and the $\tau$-regeneration. The neutral current interaction with nuclei $N$ in the Sun's medium could change the energy of neutrinos and so smears the sharp line in the spectrum; but since the cross section of neutral current scattering $\nu(\vec{k})N\to\nu(\vec{k'})X$ is finite in the limit $\vec{k'}\to\vec{k}$ (zero momentum transfer or forward scattering), the final energy spectrum of neutrinos would be a sharp line superimposed on a tail corresponding to scattered neutrinos. The $\tau$-regeneration effect, which is the reproduction of $\nu_e$ ($\bar{\nu}_e$), $\nu_\mu$ ($\bar{\nu}_\mu$) and $\bar{\nu}_\tau$ ($\nu_\tau$) from the prompt decay of $\tau^+$ ($\tau^-$) produced in the charged current interaction of $\bar{\nu}_\tau$ ($\nu_\tau$), also smears the sharp line in the spectrum. However, in the low energy range ($m_\chi \lesssim 500\, {\rm GeV}$) these effects are negligible and we focus just on the sharp line in the spectrum. A complete analysis including these effects and also annihilation modes $\chi\bar{\chi}\to b\bar{b},W^+W^-,\ldots$ will appear in \cite{future}.

In this section we consider the flavor content of the neutrino beam arrived at the Earth from the annihilation modes $\chi\bar{\chi}\to\nu_\alpha\bar{\nu}_\alpha$ where ($\alpha=e,\mu,\tau$) \footnote{recently the authors of \cite{Allahverdi:2012bi} proposed a new method to distinguish between these annihilation modes based on the $\tau$-regenerated neutrinos.}. Also we consider the annihilation $\chi\bar{\chi}\to\nu_s\bar{\nu}_s$ which is a new possibility in the presence of sterile neutrinos. Basically, the amplitude $\mathcal{A}$ of flavor oscillation $\nu_\alpha^\odot\to\nu_\beta^{{\rm Det}}$ from the center of Sun to the detector place at the Earth is given by
\begin{eqnarray}\label{amp}
\mathcal{A}(\nu_\alpha^\odot\to\nu_\beta^{{\rm Det}})=\sum_i \mathcal{A}(\nu_\alpha^\odot\to\nu_i^{\odot{\rm surf}}) \exp\left[{-i\frac{\Delta m^2_{i1}L_{\rm ES}}{2E_\nu}}\right]  \mathcal{A}(\nu_i^{\oplus{\rm surf}}\to\nu_\beta^{{\rm Det}})~,
\end{eqnarray}  
where $\nu_i^{\oplus{\rm surf}}$ represent the neutrino mass eigenstates at the surface of Earth and $\nu_\beta^{{\rm Det}}$ is the neutrino flavor state at the site of IceCub detector at South Pole. The exponential factor comes from the mass eigenstate's evolution in vacuum between the surface of Sun and the surface of Earth; and $L_{\rm ES}$ is the Earth-Sun distance $\sim 1.5\times 10^8\,{\rm km}$. The probability of flavor oscillation is given by $|\mathcal{A}(\nu_\alpha^\odot\to\nu_\beta^{{\rm Det}})|^2$. Because of the large distance $L_{\rm ES}$, the exponential factor due to $\Delta m_{41}^2$ average out from the oscillation probability; but, as it shown in \cite{farzan1,farzan3}, the exponential factors due to 12-mixing and 13-mixing do not wash out.

The neutrinos produced by the annihilation of the DM particles inside the Sun would arrive at the Earth coherently. The size of the neutrino wave-packet at the production point is determined by the mean free path $l_{\rm m.f.p.}$ of annihilating $\chi$ particles: $\sigma_x\sim l_{\rm m.f.p.} \sim 1/(n_\chi \sigma_{\rm ann})$. Assuming typical annihilation cross section determined by the DM abundance in the present Universe $\langle \sigma_{\rm ann}v \rangle\sim1\,{\rm pb}$, it is straightforward to show that $\sigma_x\sim10^{19}\,{\rm cm}$. Thus, for neutrinos with energy $E_\nu\sim100\,{\rm GeV}$, from the comparison of coherence length $L_{\rm coh}=4\sqrt{2}E_\nu^2\sigma_x/\Delta m^2\sim10^{39}\,{\rm m}/(\Delta m^2/{\rm eV}^2)$ with the Earth-Sun distance $L_{\rm ES}\sim 10^{11}$~m, it can be seen that coherence is maintained during the propagation from Sun to Earth. However, several phenomena could change the size of the neutrino wave-packet inside the Sun. It was shown in~\cite{farzan3} that among them the most important is widening of neutrino energy distribution due to the thermal velocity distribution of $\chi$ particles before annihilation, which induce the widening $\Delta E_\nu/E_\nu\sim 10^{-4}$. It is easy to see that considering this effect, still the coherence length is at least three orders of magnitude larger than the $L_{\rm ES}$ for neutrinos with the energy $E_\nu=100\,{\rm GeV}$.

Let us first consider the three active standard neutrino oscillation. It should be noticed that unlike the familiar solar neutrinos from nuclear fusion in the Sun, with the energy $E_\nu\sim {\rm MeV}$, the neutrinos from $\chi\bar{\chi}$ annihilation with energy $E_\nu\sim 100\, {\rm GeV}$ do not leave the surface of the Sun in mass eigenstates. In the case of solar neutrinos from nuclear fusion, the state of the neutrino at surface of Sun is practically $\nu_2$; but, in the case of neutrinos from $\chi\bar{\chi}$ annihilation, the state of the neutrino at the surface of Sun is a coherent superposition of $\nu_1$ and $\nu_2$ (assuming $\theta_{13}=0$). Thus, generally in the case of monochromatic neutrinos (which is the case for $\chi\bar{\chi}$ annihilation directly to neutrinos) the oscillatory terms in eq.~(\ref{amp}) do not average out. It was shown in \cite{farzan1,farzan3} that this non-averaging of oscillatory terms leads to a seasonal variation of the flavor oscillation probability due to the change in the Earth-Sun distance during the year (which is $\sim 5\times 10^6\,{\rm km}$).

\begin{figure}[t]
\centering
\subfigure[]{
\includegraphics[scale=0.42]{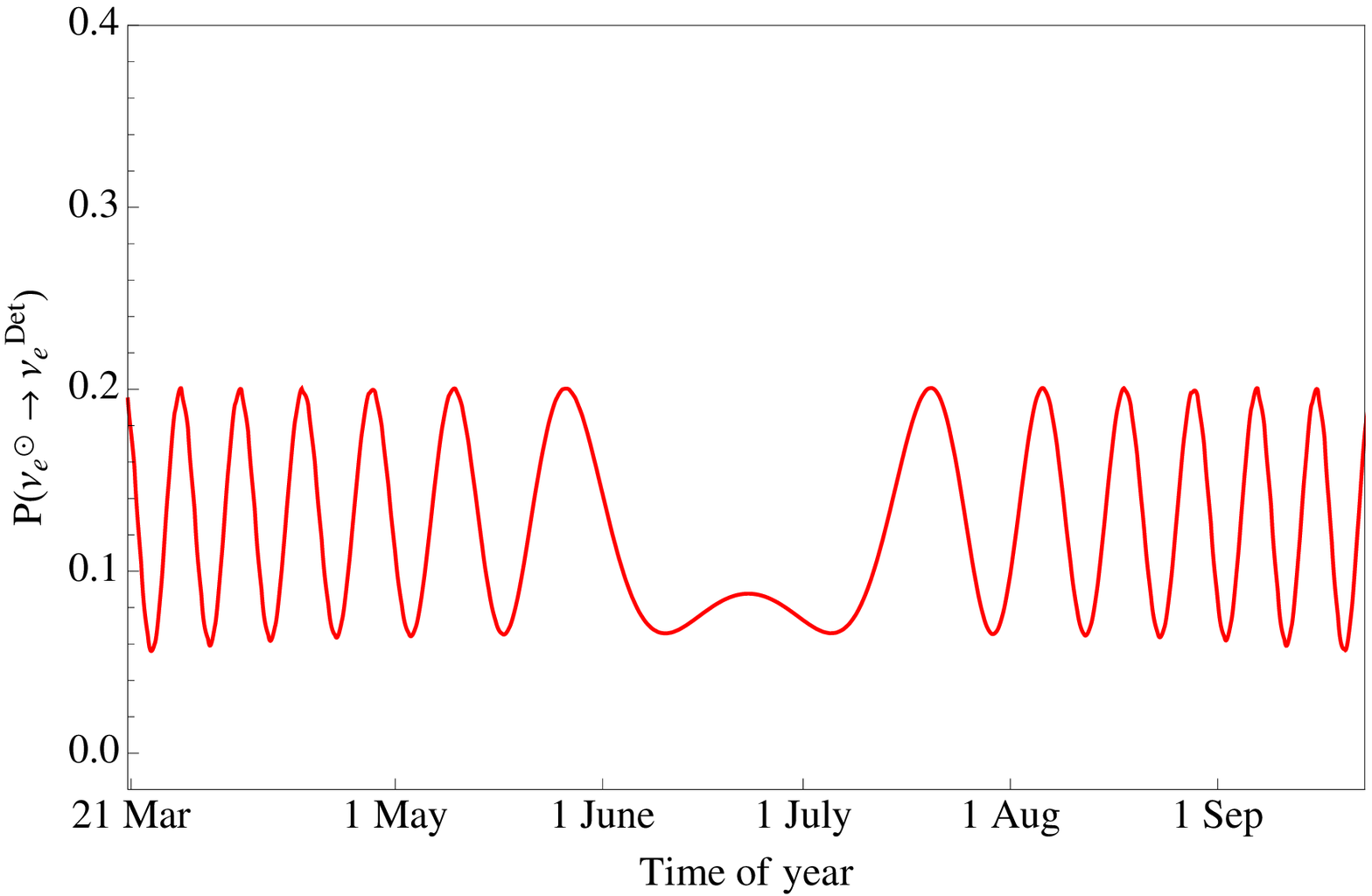}
\label{var1}
}\subfigure[]{
\includegraphics[scale=0.42]{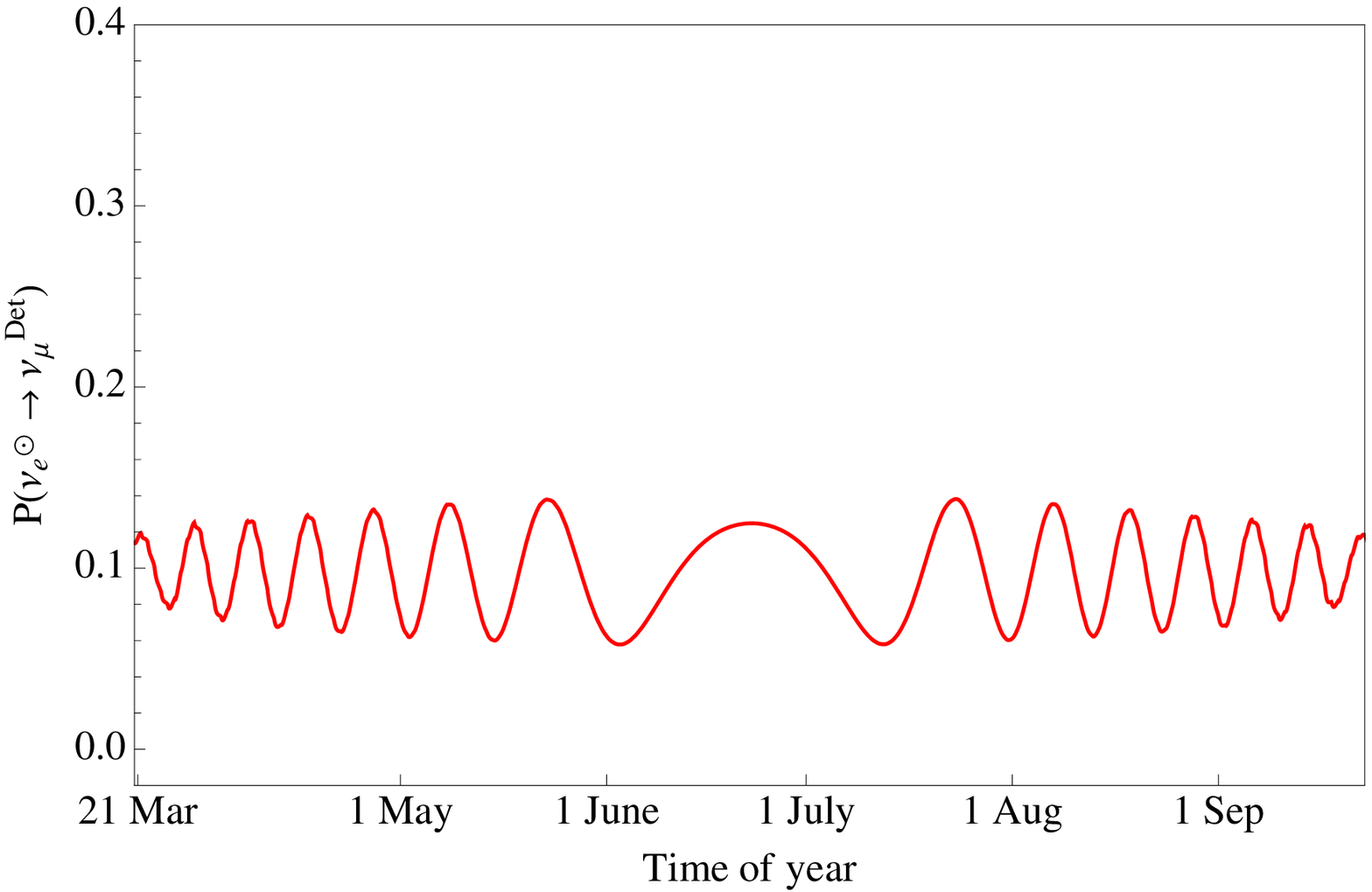}
\label{var2}
}
\caption[Optional caption for list of figures]{Seasonal variation of the probability $P(\nu_e^\odot\to\nu_e^{{\rm Det}})$ (a) and $P(\nu_e^\odot\to\nu_\mu^{{\rm Det}})$ (b). For figure (a) and (b) we assumed ($s^2_{14}=0.1,s^2_{24}=s^2_{34}=0$) and ($s^2_{14}=s^2_{24}=0.1,s^2_{34}=0$), respectively; and for both figures: $\Delta m_{41}^2=1\,{\rm eV}^2$ and $E_\nu=300\,{\rm GeV}$. }
\end{figure}

In 3+1 scenario, the MSW resonance discussed in previous section would change the seasonal variation. Let us temporarily ignore the present upper bounds on the active-sterile mixing parameters for illustrative reasons. For example, consider the case of $\chi\bar{\chi}\to\nu_e\bar{\nu}_e$ with $m_\chi>85\,{\rm GeV}$. The $\nu_e^{\odot}$ state evolving toward the surface of the Sun undergo resonance and the state of neutrino leaving the surface of Sun is mostly $\nu_4^{\odot{\rm surf}}$. Thus, the sum in eq.~(\ref{amp}) reduces to one term and taking the absolute value to obtain the probability wash out  the exponential term. However, inside the Earth, neutrinos propagate different paths toward the IceCube site during the year due to the change in Sun's zenith angle with time. We consider the period of the southern fall and winter seasons (21 March-23 September) where the Sun is below the horizon at the IceCube site and thus the background from atmospheric muons is suppressed by the Earth's matter. The zenith angle of Sun is $90^\circ$ at 21st of March and takes its maximum value at 21st June (winter solstice in southern hemisphere) which is $\sim 113.2^\circ$. As a result, the propagation length of neutrinos inside the Earth (from the surface of Earth at Sun's direction to the IceCube site) is zero at 21st March and 23rd Sep and take its maximum value $\sim 5100 \,{\rm km}$ at 21st June; and considering the $L_{{\rm osc}}^{41}$ in eq.~(\ref{14osc}), we expect to observe the oscillation pattern of 1-4 mixing at the IceCube location.  Thus, for example, in the case of $P(\nu_e^\odot\to\nu_e^{{\rm Det}})$ we expect a constant probability with respect to time which is modulated by the 14-induced oscillation inside the Earth. To illustrate this argument, we plot in figure~\ref{var1} the probability $P(\nu_e^\odot\to\nu_e^{{\rm Det}})$ with respect to time obtained by numerical solution of the evolution equations from the Sun's center to the IceCube site at South Pole.  For this figure we assumed ($s^2_{14}=0.1,\theta_{24}=0,\theta_{34}=0$), $\Delta m_{41}^2=1\,{\rm eV}^2$ and $E_\nu=300\,{\rm GeV}$. As can be seen, the pattern of the probability is a constant (with respect to time) curve modulated by an oscillatory component. The constant value is $s_{14}^2=0.12$ which comes from the decomposition of $\nu_4^{\oplus{\rm surf}}$ to $\nu_e^{\oplus{\rm surf}}$ and the oscillatory pattern comes from the 14-induces oscillation inside the Earth with the amplitude $4c_{14}^2 s_{14}^4\sim0.05$. The nearly plateau region in the middle of figure~\ref{var1} comes from this fact that at the winter solstice in southern hemisphere, the zenith angle of Sun viewed from the South Pole reach it maximum value and after this date it begins to decrease. Thus, the small rate of change in the Sun's zenith angle which leads to a small rate of change in the path length of neutrino propagation inside the Earth, suppresses the oscillation pattern at solstice dates.

Let us now consider the probability $P(\nu_e^\odot\to\nu_\mu^{{\rm Det}})$. In this case again the state of the neutrino at the surface of the Sun is almost $\nu_4^{\odot{\rm surf}}$ and from eq.~(\ref{amp}) we do not expect any seasonal variation in the probability. However, at the surface of the Earth, the $\nu_4^{\oplus{\rm surf}}$ state can be decomposed to flavor states $\nu_e^{\oplus{\rm surf}}$, $\nu_\mu^{\oplus{\rm surf}}$ and $\nu_s^{\oplus{\rm surf}}$ with coefficients $s_{14}$, $c_{14}s_{24}$ and $c_{14}c_{24}$ respectively (assuming $\theta_{34}=0$). It should be noticed that if $s_{24}=0$, the probability $P(\nu_e^\odot\to\nu_\mu^{{\rm Det}})$ would be zero; and because of this we assume $s_{24}\neq 0$ in this case. As discussed above, due to the comparable oscillation length $L_{{\rm osc}}^{14}$ with the propagation length of neutrinos inside the Earth, oscillatory pattern from $\nu_s^{\oplus{\rm surf}}\to \nu_\mu^{\rm Det}$ is expected (the oscillation probabilities of $\nu_e^{\oplus{\rm surf}}\to \nu_\mu^{\rm Det}$ and $\nu_\mu^{\oplus{\rm surf}}\to \nu_\mu^{\rm Det}$ are very small). Figure~\ref{var2} shows the seasonal variation of $P(\nu_e^\odot\to\nu_\mu^{{\rm Det}})$. In this figure we assumed $(s_{14}^2=s_{24}^2=0.1,s_{34}^2=0)$, $\Delta m_{41}^2=1\,{\rm eV}^2$ and $E_\nu=300 \,{\rm GeV}$. In this case the amplitude of oscillation is the product of $c_{14}^2s_{24}^2\sim0.1$ (from $\nu_4^{\oplus{\rm surf}}$ decomposition to $\nu_\mu^{\oplus{\rm surf}}$) by the factor $4c_{14}^4 c_{24}^2 s_{24}^2\sim0.32$ (from the oscillation probability induced by 14-mixing inside the Earth), which is smaller in comparison with $P(\nu_e^\odot\to\nu_e^{{\rm Det}})$ in figure~\ref{var1}.

The plots shown in figures~\ref{var1}~and~\ref{var2} are for illustrative purposes and the values assumed for the 3+1 mixing parameters are larger than the current upper limits. With the assumption of the best-fit values for $s_{14}^2,s_{24}^2$ and $\Delta m_{41}^2$ from the short baseline neutrino oscillation experiments \cite{giunti} and also the upper limit on $s_{34}^2$ from MINOS experiment \cite{minos}, practically the oscillation probabilities $P(\nu_e^\odot\to\nu_\mu^{{\rm Det}})$ and $P(\bar{\nu}_{\mu/\tau}^\odot\to\bar{\nu}_\mu^{{\rm Det}})$ suppress to zero. However, it should be noticed that a nonzero value for all the mixing parameters $s_{14},s_{24},s_{34}$ leads to a nonzero decomposition of $\nu_4^{\oplus{\rm surf}}$ to all the flavor states, which open new oscillation channels $\nu_\alpha^{\oplus{\rm surf}}\to \nu_\mu^{\rm Det}$ and $\bar{\nu}_\alpha^{\oplus{\rm surf}}\to \bar{\nu}_\mu^{\rm Det}$.

\begin{figure}[t]
\centering
\subfigure[]{
\includegraphics[scale=0.42]{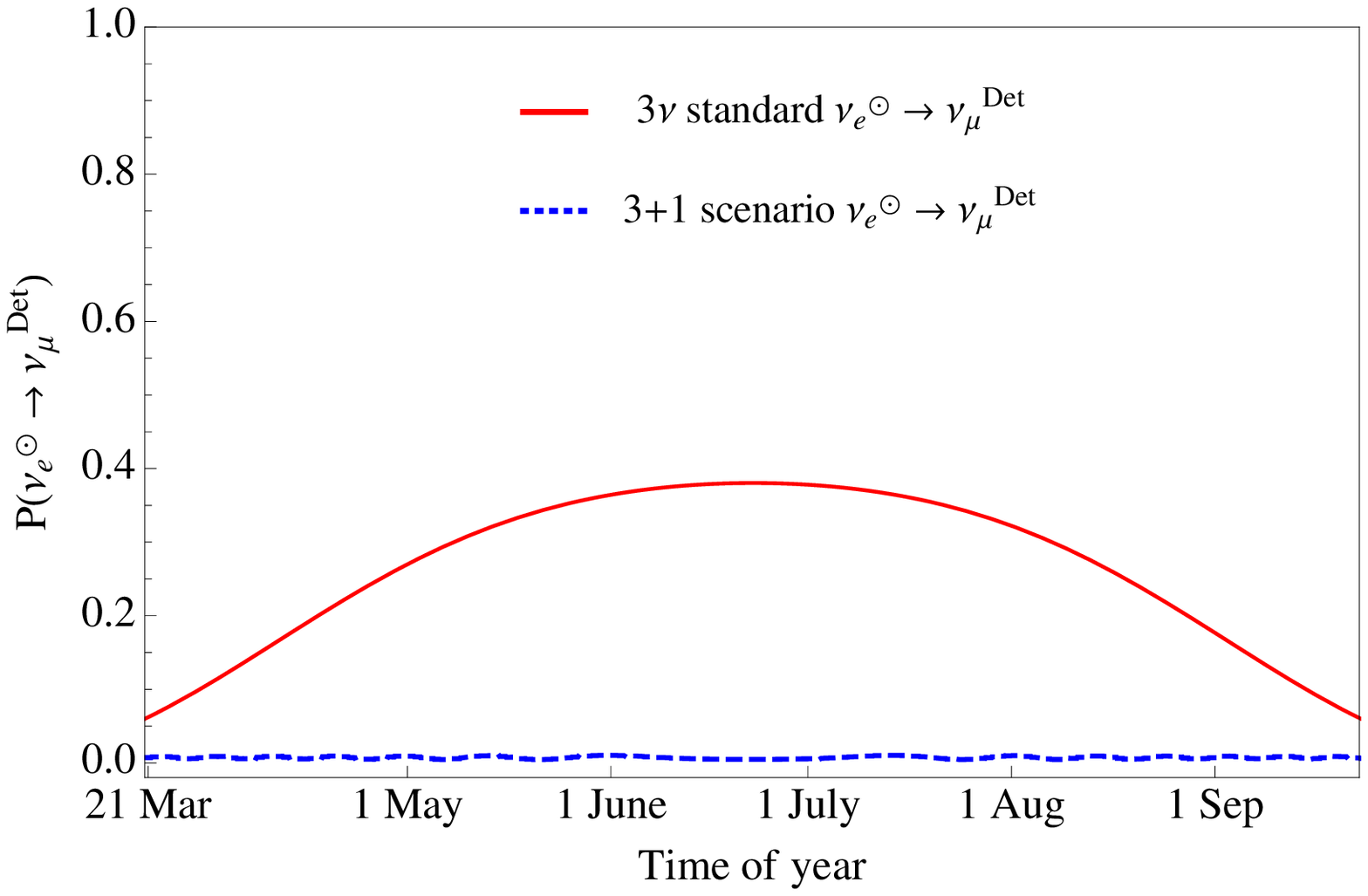}
\label{var3}
}\subfigure[]{
\includegraphics[scale=0.42]{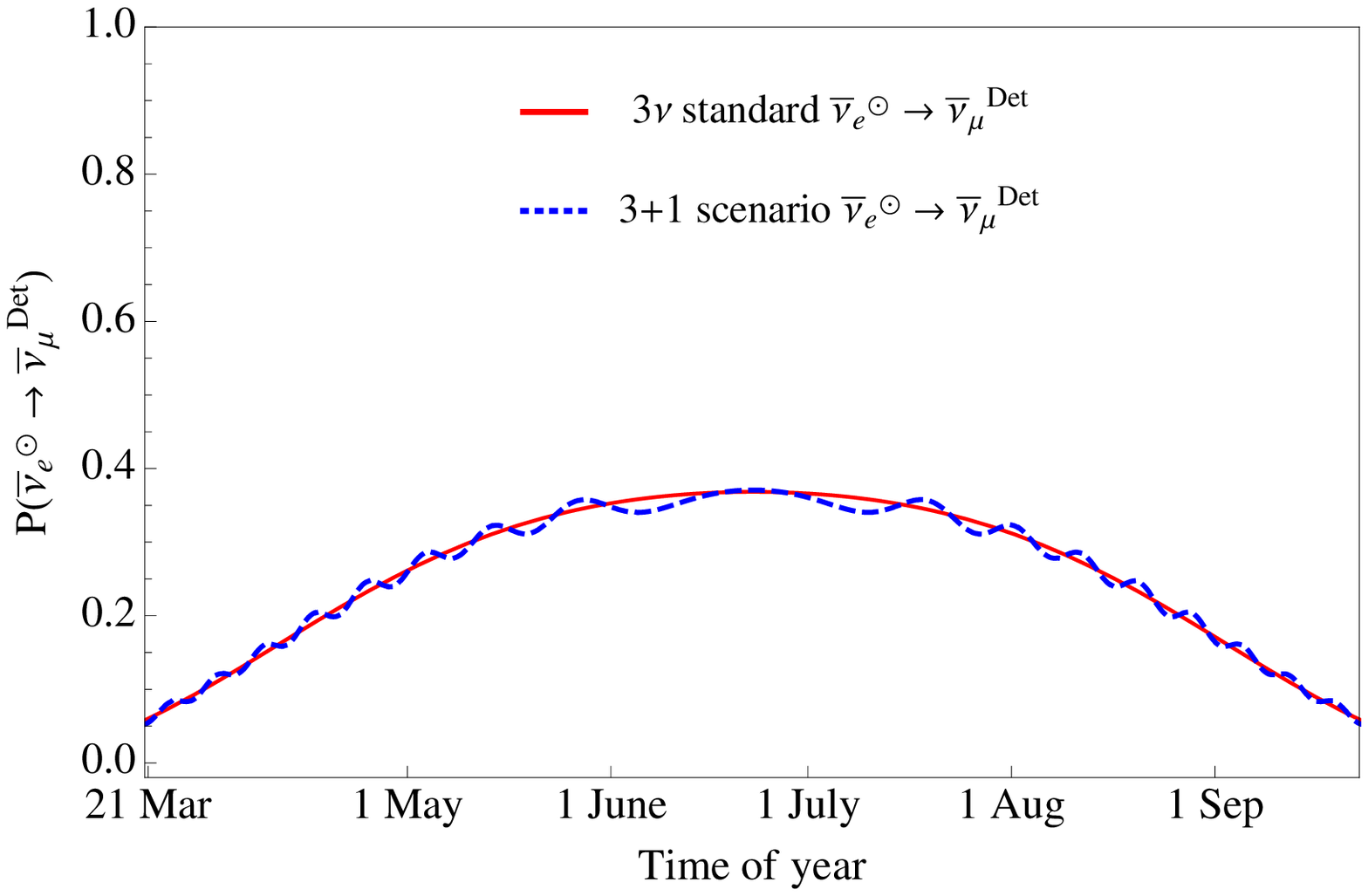}
\label{var6}
}
\caption[Optional caption for list of figures]{\label{nueprob} Seasonal variations of $P(\nu_e^\odot\to\nu_\mu^{{\rm Det}})$ (a) and $P(\bar{\nu}_e^\odot\to\bar{\nu}_\mu^{{\rm Det}})$ (b). In each plot the red (solid) curve is for 3$\nu$ standard oscillation and the blue (dashed) curve is for 3+1 scenario. In both plots we assumed $E_\nu=300\,{\rm GeV}$ and for the 3+1 scenario parameters we assumed: $s_{14}^2=s_{24}^2=s_{34}^2=0.008$ and $\Delta m_{41}^2=1\,{\rm eV}^2$. }
\end{figure}

It should be mentioned that the main detection mode of neutrinos in the energy range $\sim 300 \,{\rm GeV}$ in neutrino telescopes such as IceCube is through $\mu$-track events.  The main contribution to $\mu$-track events come from the charged current interaction of $\nu_\mu$ and $\bar{\nu}_\mu$ with the nuclei inside or in the vicinity of IceCube site. Thus, the IceCube is mainly sensitive to the $\nu_\mu$ and $\bar{\nu}_\mu$ flux at its site.

\begin{figure}[t]
\centering
\subfigure[]{
\includegraphics[scale=0.42]{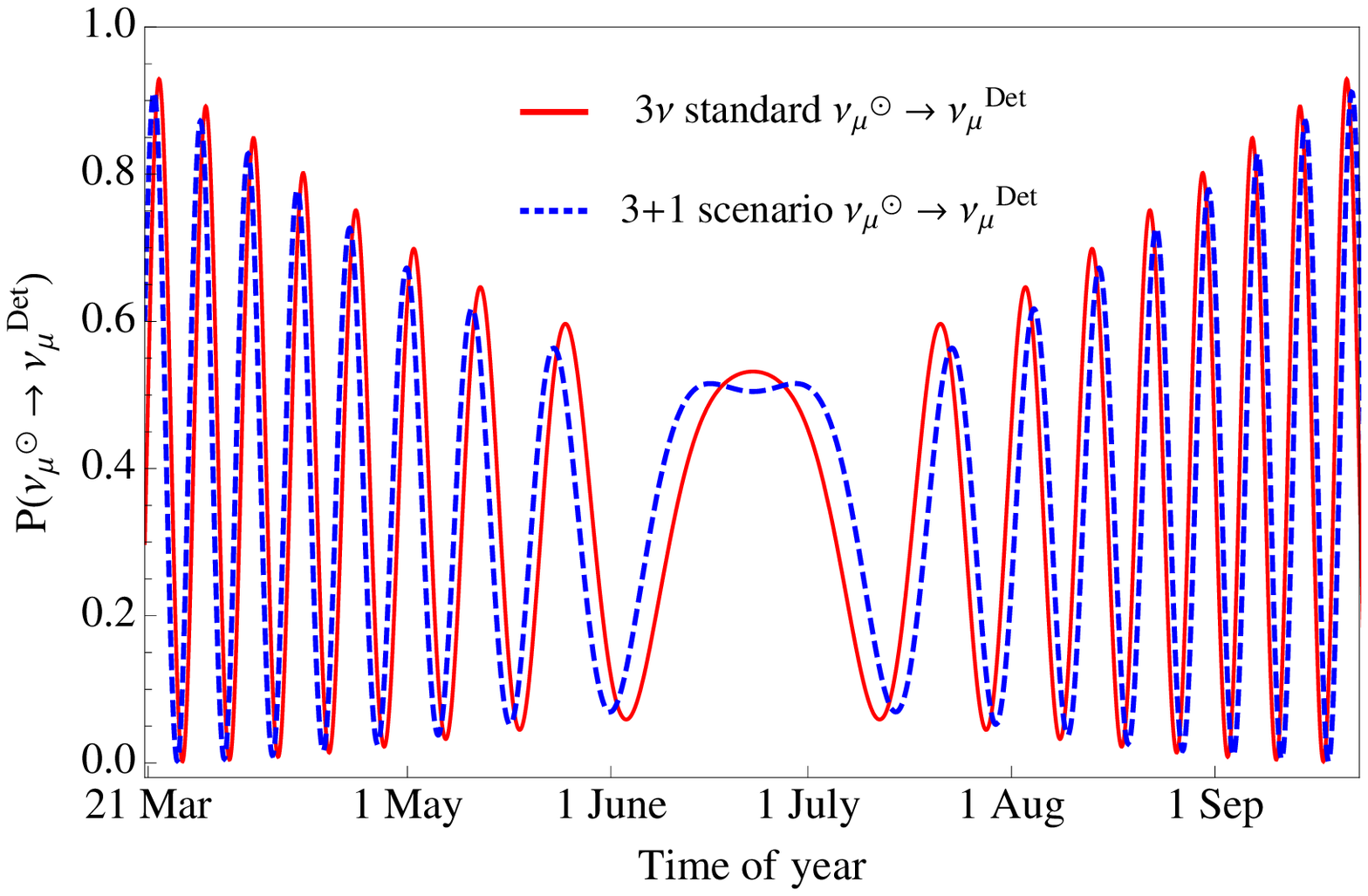}
\label{var4}
}\subfigure[]{
\includegraphics[scale=0.42]{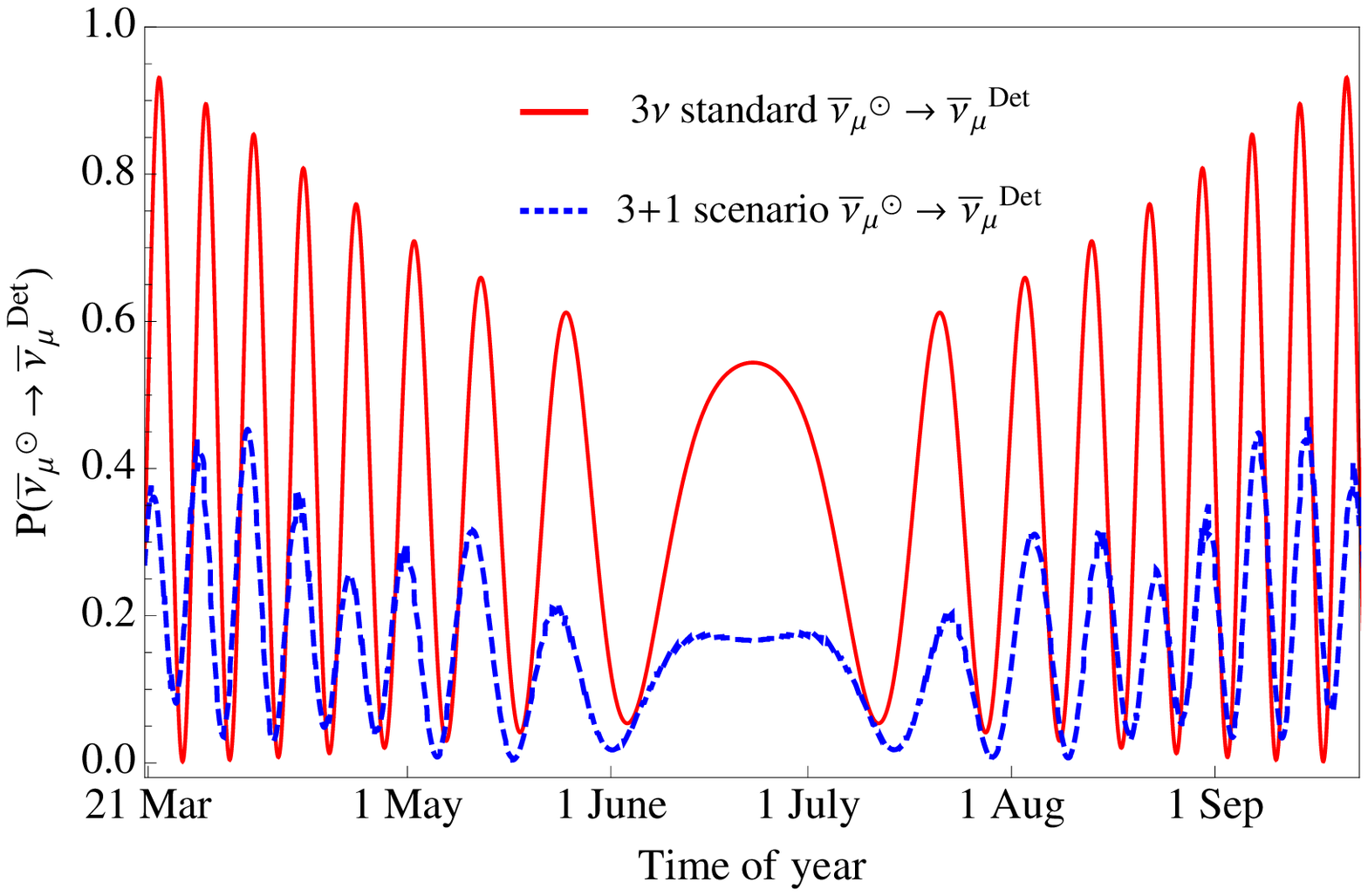}
\label{var7}
}
\caption[Optional caption for list of figures]{\label{numuprob} Seasonal variations of $P(\nu_\mu^\odot\to\nu_\mu^{{\rm Det}})$ (a) and $P(\bar{\nu}_\mu^\odot\to\bar{\nu}_\mu^{{\rm Det}})$ (b). In each plot the red (solid) curve is for 3$\nu$ standard oscillation and the blue (dashed) curve is for 3+1 scenario. In both plots we assumed $E_\nu=300\,{\rm GeV}$ and for the 3+1 scenario parameters we assumed: $s_{14}^2=s_{24}^2=s_{34}^2=0.008$ and $\Delta m_{41}^2=1\,{\rm eV}^2$. }
\end{figure}

\begin{figure}[t]
\centering
\subfigure[]{
\includegraphics[scale=0.42]{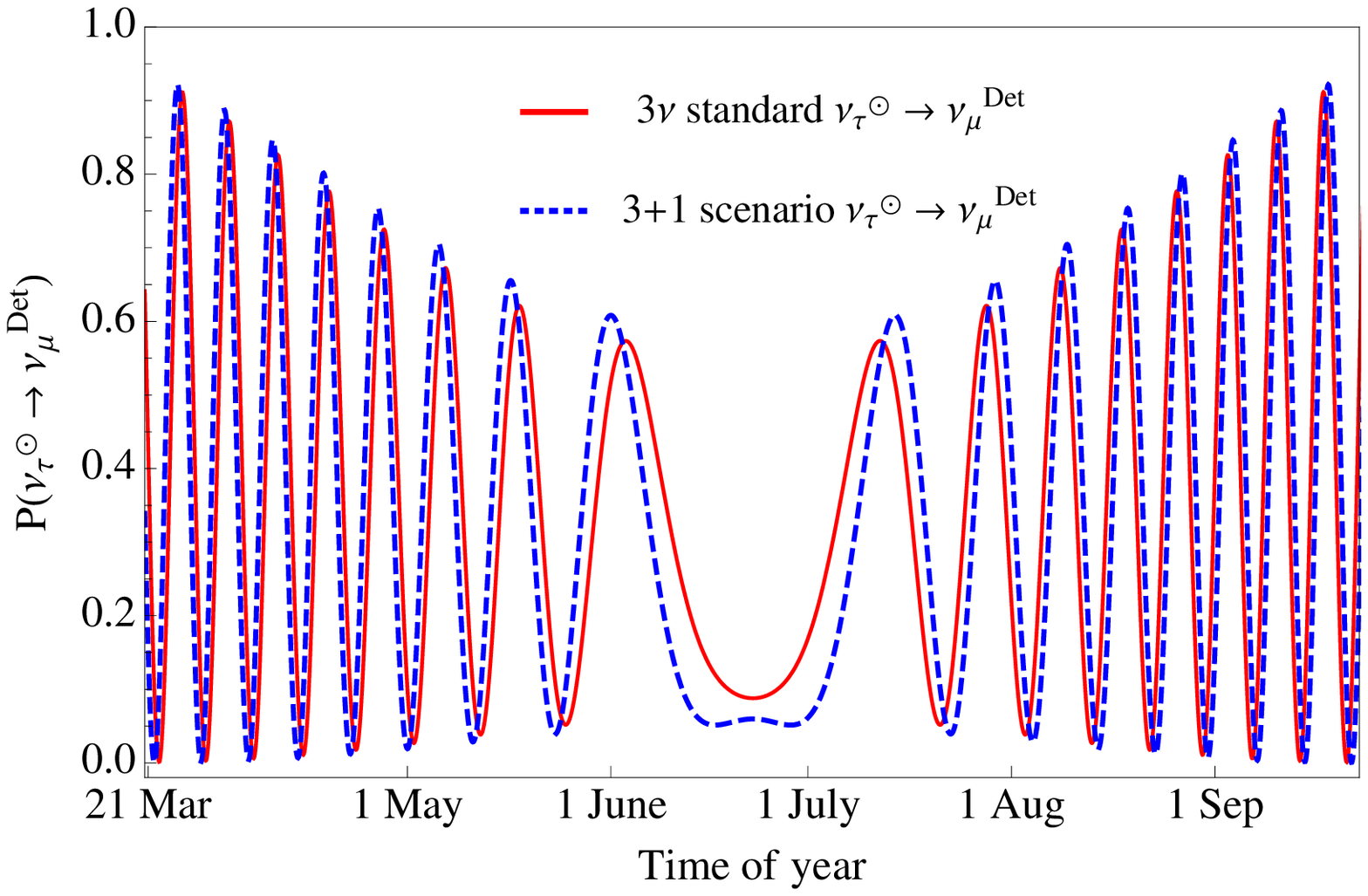}
\label{var5}
}\subfigure[]{
\includegraphics[scale=0.42]{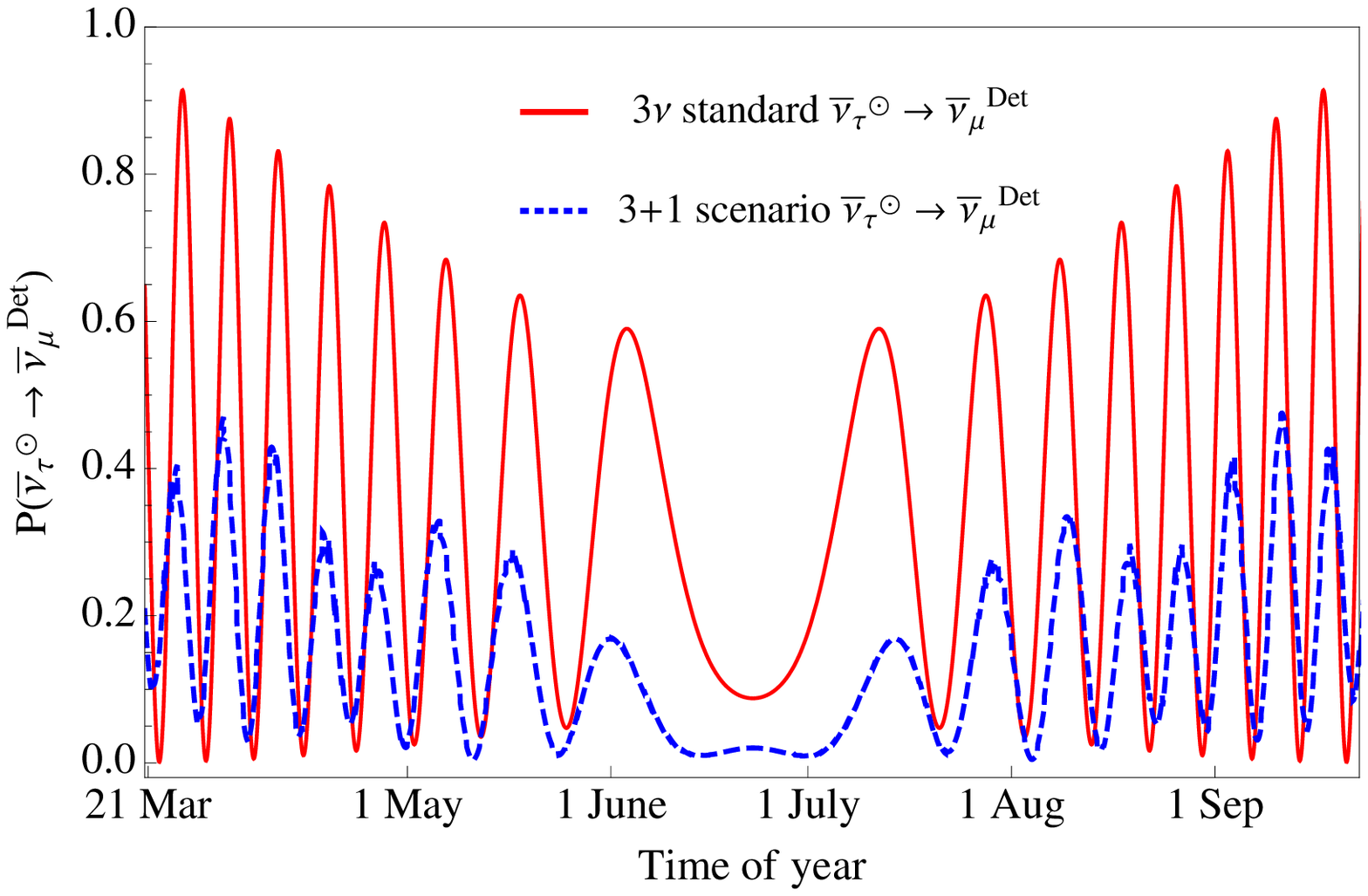}
\label{var8}
}
\caption[Optional caption for list of figures]{\label{nutauprob} Seasonal variations of $P(\nu_\tau^\odot\to\nu_\mu^{{\rm Det}})$ (a) and $P(\bar{\nu}_\tau^\odot\to\bar{\nu}_\mu^{{\rm Det}})$ (b). In each plot the red (solid) curve is for 3$\nu$ standard oscillation and the blue (dashed) curve is for 3+1 scenario. In both plots we assumed $E_\nu=300\,{\rm GeV}$ and for the 3+1 scenario parameters we assumed: $s_{14}^2=s_{24}^2=s_{34}^2=0.008$ and $\Delta m_{41}^2=1\,{\rm eV}^2$. }
\end{figure}

We have shown in figures~\ref{var3}~and~\ref{var6} the seasonal variation of $P(\nu_e^\odot\to\nu_\mu^{{\rm Det}})$ and $P(\bar{\nu}_{e}^\odot\to\bar{\nu}_\mu^{{\rm Det}})$, respectively. Figures~\ref{numuprob}~and~\ref{nutauprob} also show seasonal variation for muon and tau (anti)neutrino at the center of Sun, respectively. In each plot the blue (dashed) curve represents the seasonal variation in 3+1 scenario and the red (solid) curve corresponds to seasonal variation in 3$\nu$ standard oscillation; where in both cases $E_\nu=300\,{\rm GeV}$. For the 3+1 scenario in these plots we assumed $\Delta m_{41}^2=1\,{\rm eV}^2$ and $s_{14}^2=s_{24}^2=s_{34}^2=0.008$; which is well below the current upper limits. As can be seen, for figures~\ref{var3},~\ref{var7}~ and~\ref{var8}, where the MSW resonance occur in the Sun, the probability is suppressed; but however as we mentioned the nonzero values of $s_{24}^2$ and $s_{34}^2$ leads to a small contribution to the probability in figures~\ref{var7}~and~\ref{var8}. The oscillatory patterns in figures~\ref{var4},~\ref{var5},~\ref{var7}~and~\ref{var8} are the modulation of two components: i) the short wave-length oscillation induced by $\Delta m_{31}^2$ between the Sun and Earth; ii) the long wave-length oscillation induced by $\Delta m_{21}^2$ between the Sun and Earth. The oscillation length of 1-2 and 1-3 mixings at $E_\nu = 300\,{\rm GeV}$ are $\sim 10^7\,{\rm km}$ and $\sim 3\times10^5\,{\rm km}$, respectively. Thus, considering the change in $L_{\rm ES}$ during the southern fall and winter which is $\sim 2.5 \times 10^6 \,{\rm km}$, justifies the oscillatory patterns. It should be noticed that in the case of figures~\ref{var7}~and~\ref{var8}, the 1-4 mixing also induces oscillation inside the Earth and as can be seen in these figures the blue (dashed) curves, corresponding to 3+1 scenario, represent irregularities.

\begin{figure}[t]
\centering
\includegraphics[scale=0.47]{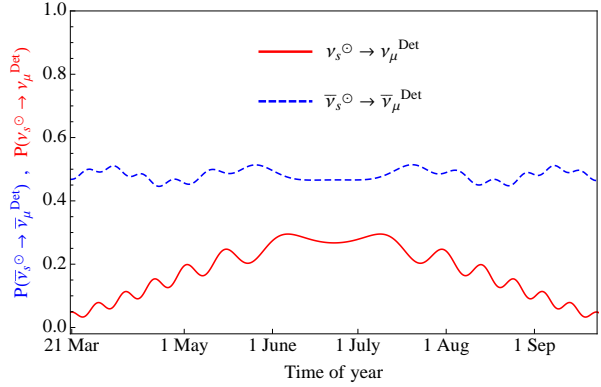}
\caption{\label{var9} Seasonal variation of the probability $P(\nu_s^\odot\to\nu_\mu^{{\rm Det}})$ (red solid curve) and $P(\bar{\nu}_s^\odot\to\bar{\nu}_\mu^{{\rm Det}})$ (blue dashed curve) for $E_\nu=400\,{\rm GeV}$. For this figure we assumed $s_{14}^2=s_{24}^2=s_{34}^2=0.008$ and $\Delta m_{41}^2=1\,{\rm eV}^2$.  }
\end{figure}

The other interesting case is the annihilation $\chi\bar{\chi}\to\nu_s\bar{\nu}_s$ in the center of Sun (for a model predicting this annihilation see \cite{farzan2}).  Figure~\ref{var9} shows the seasonal variation of $P(\nu_s^\odot\to\nu_\mu^{{\rm Det}})$ (red solid curve) and $P(\bar{\nu}_s^\odot\to\bar{\nu}_\mu^{{\rm Det}})$ (blue dashed curve) for $E_\nu=400\,{\rm GeV}$. For this figure also we assumed $\Delta m_{41}^2=1\,{\rm eV}^2$ and $s_{14}^2=s_{24}^2=s_{34}^2=0.008$. For $\nu_s^\odot$ created at the center of Sun, the state of the neutrino leaving the surface of Sun is almost $c_{12}|\nu_1^{\odot{\rm surf}}\rangle+s_{12}|\nu_2^{\odot{\rm surf}}\rangle$ and so according to eq.~(\ref{amp}) the seasonal variation due to $\Delta m_{21}^2$ is expected to be seen. The minimum and maximum of the amplitude of this oscillation with respect to time come from the decomposition of $c_{12}|\nu_1^{\odot{\rm surf}}\rangle+s_{12}|\nu_2^{\odot{\rm surf}}\rangle$ and $s_{12}|\nu_1^{\odot{\rm surf}}\rangle+c_{12}|\nu_2^{\odot{\rm surf}}\rangle$ to the state $|\nu_\mu^{\oplus{\rm surf}}\rangle$, respectively. It is straightforward to see that the maximum ($4c_{24}^2c_{23}^2c_{12}^2s_{12}^2\sim0.28$) and minimum ($s_{14}^2s_{24}^2\sim 0$) matches with the red (solid) curve in figure~\ref{var9}. For the case of $\bar{\nu}_s^\odot$, the neutrino state at the surface of Sun is mostly $|\bar{\nu}_3^{\odot{\rm surf}}\rangle$ and so we expect a constant oscillation probability $c_{24}^2s_{23}^2\sim 0.5$ (assuming $s_{13}=0$) which is in agreement with the blue (dashed) curve of figure~\ref{var9}. The small amplitude oscillation pattern in the blue (dashed) curve are due to the modulation of two components: the 14-induced oscillation inside the Earth and 13-induced oscillation from the surface of Sun to the surface of Earth which is a result of the fact that the state of the neutrino at the surface of the Sun is not exactly $|\bar{\nu}_3^{\odot{\rm surf}}\rangle$ and a few percent $|\bar{\nu}_4^{\odot{\rm surf}}\rangle$ admixture exists.

\section{\label{sec:conc}Conclusion}

We have considered the evolution of neutrinos with energy $E_\nu\sim 100 \, {\rm GeV}$ inside the Sun in the presence of new sterile neutrino state. Neutrinos in this energy range can be produced from the annihilation of DM particles ($\chi$) gravitationally trapped inside the Sun. Accumulation of DM particles in the Sun rises the number density which result in sizable DM annihilation rate. Neutrinos are the ubiquitous final product of DM annihilation either with a continuous energy spectrum from the annihilation modes $\chi\bar{\chi}\to q\bar{q},W^+W^-,\ldots$ or with a monochromatic spectrum in the direct annihilation $\chi\bar{\chi}\to \nu \bar{\nu}$. 

The recent global analysis of the data from short baseline neutrino oscillation experiments \cite{giunti} favor the presence of one (or more) mostly sterile neutrino mass eigenstates with mass splitting $\Delta m^2\sim1 \, {\rm eV}^2$. The new mass splitting $\Delta m^2\sim1 \, {\rm eV}^2$ would result in resonant flavor conversion inside the Sun. We have shown that the resonant flavor conversion would occur for $\nu_e$ with energies $E_\nu\gtrsim 85\,{\rm GeV}$ and $\bar{\nu}_{\mu/\tau}$ with energies $E_\nu\gtrsim 240\,{\rm GeV}$. This anticipated MSW flavor conversion deplete the flux of $\nu_e$ and $\bar{\nu}_{\mu/\tau}$ produced in the annihilation of DM particles. It should be noticed that the resonance and the subsequent depletion of neutrino flux take place even for very small active-sterile mixing angles, far below the present upper bounds from short baseline experiments. Thus, we propose that this point should be considered in the analysis of data from neutrino telescopes aiming at DM search from Sun. Interestingly, even non-observation of the expected excess of events from DM annihilation in the Sun's direction in neutrino telescopes can be interpreted by this phenomenon.          

As an example of this effect, we considered the case of monochromatic neutrinos produced in the direct annihilation of DM particles to neutrinos $\chi\bar{\chi}\to \nu \bar{\nu}$. It was shown in \cite{farzan1,farzan3} that for the monochromatic neutrinos, the oscillation probabilities and as a result the number of events in neutrino telescopes exhibit seasonal variation. We calculated the oscillation probabilities with the assumption of presence of sterile neutrino state. As we expected, the depletion of neutrino flux at the Earth is severe for the $\nu_e$ and $\bar{\nu}_{\mu/\tau}$ at the center of Sun. A new annihilation channel which emerge in the presence of sterile neutrinos is $\chi\bar{\chi}\to\nu_s\bar{\nu}_s$. We calculated the oscillation probabilities for this annihilation mode at the Earth. In this case, again due to the resonant flavor conversion at the Sun, the probabilities $\nu_s^\odot\to\nu_\mu^{\rm Det}$ and $\bar{\nu}_s^\odot\to\bar{\nu}_\mu^{\rm Det}$ would be significant in spite of the strong upper limits on the sterile-active mixing angles.

\acknowledgments
The authors are grateful to Y.~Farzan for her valuable comments on the manuscript. O.~L.~G.~P. thanks the Cosmology Initiative of Arizona State University, where part of this work was made, for the hospitality. A.~E. and O.~L.~G.~P.  thank support from FAPESP and O.~L.~G.~P. thanks support from CAPES/Fulbright. We acknowledge the use
of CENAPAD-SP and CCJDR computing facilities.

\end{document}